\documentstyle[11pt]{article}

\makeatletter

\if@twoside
\else
\fi
\advance\topmargin by -3mm

\ifcase \@ptsize
\or 
\or 
\fi

\def\@citex[#1]#2{%
\if@filesw \immediate \write \@auxout {\string \citation {#2}}\fi
\@tempcntb\m@ne \let\@h@ld\relax \def\@citea{}%
\@cite{%
 \@for \@citeb:=#2\do {%
 \@ifundefined {b@\@citeb}%
 {\@h@ld\@citea\@tempcntb\m@ne{\bf ?}%
  \@warning {Citation `\@citeb ' on page \thepage \space undefined}}%
 {\@tempcnta\@tempcntb \advance\@tempcnta\@ne%
 \@tempcntb\number\csname b@\@citeb \endcsname \relax%
 \ifnum\@tempcnta=\@tempcntb %
 \ifx\@h@ld\relax%
 \edef \@h@ld{\@citea\csname b@\@citeb\endcsname}%
 \else%
 \edef\@h@ld{\ifmmode{-}\else--\fi\csname b@\@citeb\endcsname}%
 \fi%
 \else
 \@h@ld\@citea\csname b@\@citeb \endcsname%
        \let\@h@ld\relax%
      \fi}%
    \def\@citea{,\penalty\@highpenalty\,}%
  }\@h@ld
}{#1}}
\@addtoreset{equation}{section}
\def\section{\@startsection {section}{1}{\z@}{-3.5ex plus -1ex minus
 -.2ex}{2.3ex plus .2ex}{\large\bf\centering}}
\def\subsection{\@startsection{subsection}{2}{\z@}{-3.25ex plus%
 -1ex minus -.2ex}{1.5ex plus .2ex}{\sc}}
\gdef\@publabel{\hfil}
\gdef\@pubdate{\null}
\gdef\@pubnumber{\null}
\gdef\@author{\null}
\gdef\@title{\null}
\gdef\@abstract{\null}
\long\def\pubdate#1{\gdef\@pubdate{#1}}
\long\def\pubnumber#1{\gdef\@pubnumber{#1}}
\long\def\publabel#1{\gdef\@publabel{#1}}
\long\def\author#1{\gdef\@author{#1}}
\long\def\title#1{\gdef\@title{#1}}
\long\def\abstract#1{\gdef\@abstract{#1}}
\def\titlerelax{
\let\maketitle\relax
\let\settitleparameters\relax
\let\consolidatetitle\relax
\let\inittitlepage\relax
\let\finishtitlepage\relax
\let\titlepagecontents\relax
\let\multithanks\relax
\let\titlebaselines\relax
\let\@makepub\relax
\let\@maketitle\relax
\let\@makeauthor\relax
\let\@makeabstract\relax
\let\@maketitlenote\relax
\let\thanks\relax
\let\titlerelax\relax}
\def\titleclean
{\gdef\@titlenote{}
\gdef\@abstract{}
\gdef\@author{}
\gdef\@title{}
\gdef\@pubdate{}\gdef\@pubnumber{}\gdef\@publabel{}
\gdef\@dpublabel{}
}

\def\@makepub{\vbox to \z@{\hbox to \textwidth{\hfill
\@publabel \hfill
\llap{\parbox[t]{0.33\textwidth}{\raggedleft\@pubnumber}}}%
\vss}}
\def\@maketitle{\vskip 60pt \begin{center}
 {\LARGE \@title \par}
 \end{center}}

\def\@makeauthor{{%
\def\and{\smallskip {\normalsize \rm and\smallskip }}
\def\And{\medskip {\normalsize \rm and\\}\medskip}
\long\def\address##1{{\def\and{\\and\\}\medskip
				{\small \it \\##1\\}
}}
{\centering
 \vskip 3em
 \large \lineskip .75em
 \@author}
 \par}}
\def\@makedate{\vskip 1.5em
 \rightline{\raggedright \small \noindent\@pubdate}}
\def\@makeabstract{\vskip 1.5em
{\small
\begin{center}
{\bf ABSTRACT\vspace{-.5em}\vspace{0pt}}
\end{center}
\quotation \@abstract \endquotation}}
\def\maketitle{\titlepage
\let\footnotesize\small \setcounter{page}{0}
\@makepub
\@makedate
\vfil
\@maketitle
\@makeauthor
\vfil
\@makeabstract
\@thanks
\vfil
\titlerelax \titleclean
\setcounter{footnote}{0}
}

\def\bigans{y }

\read-1 to\answ
\ifx\answ\bigans
\read-1 to\bnsw
	\ifx\bnsw\bigans 

\ifcase\@ptsize
 \font\tenmsa=msam10
 \font\sevenmsa=msam7
 \font\fivemsa=msam5
 \font\tenmsb=msbm10
 \font\sevenmsb=msbm7
 \font\fivemsb=msbm5
\or
 \font\tenmsa=msam10 scaled \magstephalf
 \font\sevenmsa=msam8
 \font\fivemsa=msam6
 \font\tenmsb=msbm10 scaled \magstephalf
 \font\sevenmsb=msbm8
 \font\fivemsb=msbm6
\or
 \font\tenmsa=msam10 scaled \magstep1
 \font\sevenmsa=msam8
 \font\fivemsa=msam6
 \font\tenmsb=msbm10 scaled \magstep1
 \font\sevenmsb=msbm8
 \font\fivemsb=msbm6
\fi

\newfam\msafam
\newfam\msbfam
\textfont\msafam=\tenmsa  \scriptfont\msafam=\sevenmsa
  \scriptscriptfont\msafam=\fivemsa
\textfont\msbfam=\tenmsb  \scriptfont\msbfam=\sevenmsb
  \scriptscriptfont\msbfam=\fivemsb

\def\hexnumber@#1{\ifnum#1<10 \number#1\else
 \ifnum#1=10 A\else\ifnum#1=11 B\else\ifnum#1=12 C\else
 \ifnum#1=13 D\else\ifnum#1=14 E\else\ifnum#1=15 F\fi\fi\fi\fi\fi\fi\fi}

\def\msa@{\hexnumber@\msafam}
\def\msb@{\hexnumber@\msbfam}

\def\Bbb{\ifmmode\let\next\Bbb@\else
 \def\next{\errmessage{Use \string\Bbb\space only in math mode}}\fi\next}
\def\Bbb@#1{{\Bbb@@{#1}}}
\def\Bbb@@#1{\fam\msbfam#1}
	\else
		\ifx\cnsw\bigans 

\ifcase\@ptsize
 \font\tenmsx=msxm10
 \font\sevenmsx=msxm7
 \font\fivemsx=msxm5
 \font\tenmsy=msym10
 \font\sevenmsy=msym7
 \font\fivemsy=msym5
\or
 \font\tenmsx=msxm10 scaled \magstephalf
 \font\sevenmsx=msxm8
 \font\fivemsx=msxm6
 \font\tenmsy=msym10 scaled \magstephalf
 \font\sevenmsy=msym8
 \font\fivemsy=msym6
\or
 \font\tenmsx=msxm10 scaled \magstep1
 \font\sevenmsx=msxm8
 \font\fivemsx=msxm6
 \font\tenmsy=msym10 scaled \magstep1
 \font\sevenmsy=msym8
 \font\fivemsy=msym6
\fi

\newfam\msxfam
\newfam\msyfam
\textfont\msxfam=\tenmsx  \scriptfont\msxfam=\sevenmsx
  \scriptscriptfont\msxfam=\fivemsx
\textfont\msyfam=\tenmsy  \scriptfont\msyfam=\sevenmsy
  \scriptscriptfont\msyfam=\fivemsy

\def\hexnumber@#1{\ifnum#1<10 \number#1\else
 \ifnum#1=10 A\else\ifnum#1=11 B\else\ifnum#1=12 C\else
 \ifnum#1=13 D\else\ifnum#1=14 E\else\ifnum#1=15 F\fi\fi\fi\fi\fi\fi\fi}

\def\msx@{\hexnumber@\msxfam}
\def\msy@{\hexnumber@\msyfam}

\def\Bbb{\ifmmode\let\next\Bbb@\else
 \def\next{\errmessage{Use \string\Bbb\space only in math mode}}\fi\next}
\def\Bbb@#1{{\Bbb@@{#1}}}
\def\Bbb@@#1{\fam\msyfam#1}

\else
\def\Bbb#1{{\bf #1}}
		\fi
	\fi
\else
\def\Bbb#1{{\bf #1}}
\fi

\begin{document}

\pubdate{}
\pubnumber{DAMTP-93-48\\ EFI-93-54\\ hep-th/9309146\\21 September, 1993 }
\title{Null vectors, 3-point and 4-point functions in\\
 conformal field theory\footnote{Talk presented by GMTW at the
III International conference on Mathematical Physics,
String Theory and Quantum Gravity, Alushta, Ukraine}%
}

\author{
P.~Bowcock%
\address{Enrico Fermi Institute,
University of Chicago,
Chicago, IL 60637, U.S.A.\thanks{Address from 1 Oct: Dept.\ Math.\ Sci.\,
Univ.\ Durham, South
Road, Durham, DH1 3LE, U.\ K.}
}
\And
G.\ M.\ T.\ Watts%
\address{St.\ John's College, St.\ John's Street, Cambridge, CB2 1TP,
U.\ K.
\\ and \\
DAMTP, University of Cambridge, Silver Street, Cambridge, CB3 9EW,
U.\ K.}%
}

\def\vec#1{|#1\rangle}
\def\cev#1{\langle#1|}
\def\eq{\begin{equation}}
\def\en{\end{equation}}
\def\eqq{\begin{eqnarray}}
\def\enn{\end{eqnarray}}

\abstract{
We consider 3-point and 4-point  correlation functions in a conformal field
theory with a W-algebra symmetry.
Whereas in a theory with only Virasoro symmetry the three point
functions of descendants fields are uniquely
determined by the three point function of the corresponding primary
fields this is not the case for a theory with $W_3$ algebra symmetry.
The  generic
3-point functions of W-descendant fields have a countable degree of
arbitrariness.
We find, however, that if one of the fields belongs to a
representation with null states that this has implications for the
3-point functions. In particular if one of the representations is
doubly-degenerate then the 3-point function is determined
up to an overall constant.
We extend our analysis to 4-point functions and find that if two of the
W-primary fields are doubly degenerate then the intermediate
channels are limited to a finite set and that the corresponding chiral
blocks are determined up to an overall constant. This corresponds to
the existence of a linear differential equation for the chiral blocks
with two completely degenerate fields as has been found in the work of
Bajnok~et~al.
}

\maketitle

\noindent%
In this talk I would like to discuss some features of correlation
functions in two-dimensional conformal field theory with extended
conformal symmetry algebras, and show that
there are some new and interesting results in this area.
In the first half I  shall review the theory in the case where the
chiral algebra is the Virasoro algebra and subsequently discuss the
$W_3$ algebra.
This analysis was inspired by the results on $W$-algebra correlation
functions  of Bajnok et al.~\cite{Bajnetal}

\setcounter{section}{1}
\setcounter{subsection}{-1}

\subsection{Virasoro algebra representation theory}

The Virasoro algebra has generators $L_m$ with commutation relations
\eq
[L_m,L_n] = \frac c{12}m(m^2-1)\delta_{m+n,0} + (m-n)L_{m+n}
\;.
\en
There are two sorts of representations of interest.

Highest weight representations have the spectrum of $L_0$ bounded from
below. The Verma module $M_{h,c}$ is a highest weight representation
and is spanned by the states
\eq
L_{i_1} \ldots L_{i_k} \vec h
,\qquad
i_j \leq i_{j+1} < 0,
\label{eq.b1}
\en
where $\vec h$ satisfies
\eq
L_m \vec h = h\delta_{m,0} \vec h\;,\qquad m\ge 0
\;.
\en
For generic $h$, $c$, $M_{h,c}$ is irreducible. If however $h$ and $c$
satisfy
\eq
c=c(t)=13 - 6t - 6/t,\;\;
h=h_{r,s}(t) = \frac{ (rt - s)^2 - (t-1)^2 }{4t},
\label{eq.vp}
\en
for some values of $t \in\Bbb R$, $r,s\in \Bbb N$,
then $M_{h_{r,s}(t),c(t)}$ is reducible with an invariant submodule
generated by a highest weight state $N_{r,s}\vec{h_{r,s}}$ at level
$rs$ (see ref.\ \cite{Kac2}).
The Virasoro minimal models have $c$ given by eq.\ (\ref{eq.vp}) with
$t=\sqrt{p/q}$, $p,q\in \Bbb N$, $p,q$ coprime, and the fields in the
models have
$h=h_{r,s}$ with $r,s\in\Bbb Z$, $0<r<q$, $0<s<p$  (see
\cite{BPZ}).

If $M_{h,c}$ is reducible than
the irreducible highest weight representation $L_{h,c}$  is the
quotient of $M_{h,c}$ by its maximal invariant submodule, which is
the union of Verma modules generated by embedded highest weight states
\cite{FeFu2}.
For physical reasons one expects the Hilbert space of a theory to
be composed of irreducible representations and that the null vectors
decouple from all correlation functions.

There are also field representations, of which the
simplest is a primary field\footnote{This representation is
the same as the $F_{\lambda,\mu}$ in \cite{FeFu1}}.
A primary field $\phi_h^{(y)}$ is labelled by $h$ and $y$
and satisfies
\eq
[ L_m, \phi^{(y)}_h(z) ] = z^{m+1}\partial\phi_h^{(y)}(z)
                         + h(m+1)z^m\phi_h^{(y)}(z)
\en
\eq
\phi_h^{(y)}(z) = \sum_{n\in \Bbb Z} z^{y-n} \phi(h,y)_n
\en
although we shall often drop the dependence on $y$ when it is clear
from context.
Heuristically a field $\Phi(z)$ satisfies
\eq
 \Phi(z) \vec 0 = e^{z L_{-1} } \vec\Phi
\en
where $\vec \Phi$ is a state in a highest weight representation. A
primary field $\phi_h$ corresponds to $\vec\Phi=\vec h$ being a
highest weight state.

\subsection{3-point functions}

In ref.\ \cite{FeFu3} Feigin and Fuchs showed us how to consider
three-point functions as a map from a Verma module
$M_{h,c}$ to $\Bbb C$,
\eq
\vec\psi \mapsto \cev{h_2} \phi_{h_3}^{(y)}(z) \vec\psi
,\;\;
y={(h_2 - h_3 - h_1)},\;\;
\psi \in M_{h,c},
\label{eq.map}
\en
\eq
\cev{h_2} \phi_{h_3}(z) \vec h = C z^{h_2 - h_3 - h_1}
\nonumber
\en
where $C$ is an arbitrary constant and $z$ a non-zero complex number.
Following ref.\ \cite{FeFu3}, we can easily evaluate (\ref{eq.map}) by
considering the following combinations of generators,
\eq
e_m(z) = L_m - 2z L_{m-1} + z^2L_{m-2},\;\;
e_0'(z) = z^2 L_{-2} - 2z L_{-1} + L_{0},\;\;
e_0''(z) = z^2 L_{-2} - z L_{-1} .
\en
These have the property that
\eq
\cev{h_2} \phi_{h_3}(z) e_m(z) = 0,\; m \leq -1,\;\;
\cev{h_2} \phi_{h_3}(z) (e_0'(z) - h_2)=0,\;\;
\cev{h_2} \phi_{h_3}(z) (e_0''(z) - h_3)=0.
\en
We shall often consider the case $z=1$ in which case we drop the $z$
dependence in the generators $e_0'$, $e_0''$, $e_m$.

The Verma module $M_{h,c}$ can be spanned by states of
the form
\eq
(e_0')^a (e_0'')^b e_{i_1}\ldots e_{i_j} \vec h
,\;\;
i_k \leq i_{k+1} \leq -1
\,.
\label{eq.ebasis}
\en
As a consequence, any state $\vec\psi$ in $M_{h,c}$ can be written in
the form
\eq
\vec\psi = p(e_0', e_0''; e_{-1}, e_{-2}, \ldots) \vec h
,
\en
where $p$ is a polynomial. The three point functions of this state is
very easy to calculate,
\eq
\cev{h_2} \phi_{h_3}(1)\vec\psi =
 p(h_2, h_3; 0,0, \ldots)  C .
\label{eq.pf}
\en

\subsection{3-point functions and null vectors}

As stated above, Feigin and Fuchs defined a three point function
initially on a Verma module. If this Verma module is reducible, then we
must check whether we can restrict this map to the irreducible
representation, that is that whether the null vectors decouple.
The null vectors can of course also be written in the form
(\ref{eq.ebasis}),
\eq
N_{p,q} = N_{p,q}(e_0',e_0'';e_{-1},\ldots) ,
\en
and so from eqn.\ (\ref{eq.pf})
\eq
\cev{h_2} \phi_{h_3}(1) N_{p,q}\vec{h_{p,q}} =
N_{p,q}(h_2, h_3; 0,0, \ldots)  C .
\en
The requirement that the 3-point function can be restricted from the
Verma module to the irreducible representation is that the polynomials
$N_{p,q}(h_2, h_3; 0,0, \ldots)$ vanish for each highest weight state
$N_{p,q}\vec{h_{p,q}}$ in $M_{h,c}$. If there is a single such highest
weight states in $M_{h,c}$ then if $h_3$ is given, $h_2$
is restricted to a finite set by this requirement.
If there is an additional second independent null vector as is the
case for the representations arising in the minimal models
then this will further restrict the allowed fusions and
for those representations which occur in the minimal models the
fusions are restricted to a finite set of allowed pairs $\{h_2,h_3\}$
(see \cite{FeFu3}).
The polynomials $N_{p,q}(h_2, h_3; 0,0, \ldots) $ have been found for
all $p,q$ indirectly by Feigin and Fuchs in ref.\ \cite{FeFu3},
directly for $p=1$ or $q=1$ by Langlands in \cite{Lang1} and by Bauer
et al.\ in \cite{BDIZ} and directly by Kent for all $p,q$ in
\cite{Kent4}. These polynomials  give the usual fusion rules for the
Virasoro algebra as predicted in \cite{BPZ}.

\subsection{4-point functions}

A four point function of primary fields is heuristically something of
the form
\eq
\cev{h_2} \phi_{h_3}(1) \phi_{h_4}(z) \vec{h_1}
\en
A chiral block is given by projecting onto a single intermediate
representation, $L_{h_5,c}$. We can of course write this projector
using an orthonormal basis $\vec{\psi_\iota}$ for $L_{h_5,c}$,
\eq
\pi = \sum_\iota \vec{\psi_\iota}\cev{\psi_\iota}
\en
The chiral block $B(h_2,h_3,h_4,h_1;h_5)$ is then given as
\eq
B(h_2,h_3,h_4,h_1;h_5) =
\sum_\iota
\cev{h_2} \phi_{h_3}(1)\vec{\psi_\iota}
\cev{\psi_\iota}  \phi_{h_4}(z) \vec{h_1}
\en
Since the 3-point functions in this chiral block are determined up to
an overall constant, we find that the chiral block is itself determined
by $\{h_i,c\}$ up to an overall scale.
Let us consider the consequences if one of the Verma modules
$M_{h_i,c}$, $i=1,2,3,4$, is reducible.  Without loss of generality,
let $h_1 = h_{p,q}(t), c=c(t)$ so that there is a null state in
$M_{h_1,c}$, which must decouple
from the correlation function. This means that the 3-point function
\eq
\cev{h_5} \phi_{h_4}(z) N_{p,q}\vec{h_{p,q}}
\en
must vanish and so we obtain the restriction that $h_5$ lies in a
finite set. Since the 4-point function is given as a sum over chiral
blocks, we see that the 4-point function lies in a finite-dimensional
space, and can be given as the solution of a linear ordinary
differential equation in $z$.
This is of course the usual way that the 4-point functions are
presented, as solutions of the differential equations expressing the
decoupling of the null vector,
as in ref.\ \cite{BPZ}; we present the results this way in order to contrast
them with the corresponding results for the $W_3$ algebra.

\setcounter{section}{2}
\setcounter{subsection}{-1}
\subsection{$W_3$ Algebra representation theory}

The $W_3$ algebra was introduced by Zamolodchikov in ref.\
\cite{Zamo1} and is the simplest
of the extended algebras which still displays interesting properties.
It has generators $L_m, Q_m$ with commutation relations
\begin{eqnarray}
{}~[ L_m, Q_n ] &=& (2m-n)Q_{m+n} \nonumber\\
{}~[ Q_m, Q_n ] &=&
\frac{(22 + 5c)}{48}\frac{c}{3\cdot 5!} (m^2-4)(m^2-1)m\delta_{m+n}
\\
&& + \frac{1}{3}(m-n)\Lambda_{m+n} +
\frac{(22 + 5c)}{48}\frac{(m-n)}{30}(2m^2-mn+2n^2-8)L_{m+n}
\,,
\nonumber
\end{eqnarray}
where
\eq
\Lambda_m = \sum_{p>-2} L_{m-p}L_p
	+   \sum_{p\leq -2} L_p L_{m-p}
	-   \frac3{10}(m+2)(m+3)L_m
,
\en
and $c$ is a central element.
The fact that this is not a Lie algebra leads to many interesting
problems.

The representation theory of the $W_3$ algebra can be developed in
analogy with that of the Virasoro algebra. A $W_3$ highest weight
vector $\vec{h,q}$ satisfies
\eq
L_m\vec{h,q}=\delta_{m,0}h\vec{h,q}\;,\;
Q_m\vec{h,q}=\delta_{m,0}q\vec{h,q}\;,\;
m \ge 0\,.
\en
The Verma module  $M_{h,q,c}$ of the $W_3$-algebra is spanned by states
of the form
\eq
L_{i_1} \ldots L_{i_j} Q_{k_1} \ldots Q_{k_l} \vec{h,q},\;\;
i_m \leq i_{m+1} \leq -1,\;
k_m \leq k_{m+1} \leq -1,
\label{eq.qbasis}
\en
and by the usual abuse of notation the central element $c$ takes the
value $c$.
If the Verma module is reducible, then the irreducible representation
$L_{h,q,c}$ is the quotient of the Verma module by its maximal
invariant submodule (note this submodule does not have to be the union
of Verma modules
generated by highest weight states, see e.g. ref.\ \cite{Watt1}).

We can parameterise the weights
of a W-highest weight vector as follows \cite{FZam4},
\eq
h=\frac{1}{3}( x^2 + xy + y^2 - 3 a^2)
\;,\; w
 = \frac{1}{27}(x-y)(2x+y)(x+2y)
\,,
\label{eq.weig}
\en
where we define $a,\alpha_\pm$ by
\eq
c= 2 - 24 a^2 \;,\; \alpha_\pm^2 - \alpha_\pm a - 1 = 0
\,.
\label{eq.cval}
\en
The condition that $M_{h,q,c}$ has a null vector with eigenvalues
$h',q'$ is that we can find some $x, y$ such that $h, q$ are given
by eqns.\ (\ref{eq.weig}) and $x$ satisfies
\eq
x = r\alpha_+  + s\alpha_-,
\;\; r,s\in \Bbb N,\;\; rs>0,
\en
in which case $h',q'$ are given by eqns.\ (\ref{eq.weig}) with
$x' = x - 2 r\alpha_+$, $y'= y + r\alpha_+$.
The $W_3$ minimal models are those which have $\alpha_+ = \sqrt{p/q}$,
$p,q\in\Bbb N$, $p,q$ coprime, and the fields in these models have
$x=r\alpha_+ + s\alpha_-$, $y=t\alpha_+ + u\alpha_-$,
$0<r,s,t,u$, $r+t<q$, $s+u<p$
(see \cite{Fluk}). Of these minimal models those with $p=m+1$, $q=m$,
$m\geq 3$ are unitary since they can be constructed in the explicitly
unitary coset construction \cite{BBSS}.

The highest weight representation theory of the $W_3$
algebra is analogous to that of the Virasoro and the Kac-Moody algebras;
primary fields of the $W_3$ algebra are not as easy to
characterise as those of the Virasoro however.

If we denote the field
corresponding to the state $X_{-n}\vec\Phi$ by $\hat X_{-n} \Phi(z)$,
then a $W_3$-primary field corresponding to a highest weight state
$\vec{h,q}$ should satisfy
\eq
{}~[Q_m, \phi_{h,q}(z)]
=
z^m \left\{ \frac 12 q(m+1)(m+2) + (m+2) z \hat Q_{-1}
	+ z^2 \hat Q_{-2} \right\} \phi_{h,q}(z)
,
\en
where $\hat Q_{-1}\phi, \hat Q_{-2}\phi$ are new fields. For generic
values of $h$, $q$  and $c$ it is not
possible to find a finite set of fields which closes under the action
of the $W_3$ algebra.
However, we can again use the trick of Feigin and Fuchs and introduce
the combinations of generators
\begin{eqnarray}
f_m(z) &=& Q_m - 3zQ_{m-1} + 3 z^2 Q_{m-2} - z^3 Q_{m-3} ,\;\;
m \leq -1 \nonumber\\
f_0'(z) &=& Q_0 - 3zQ_{-1} + 3 z^2 Q_{-2} - z^3 Q_{-3} \\
f_0''(z) &=& -z Q_{-1} + 2z^2 Q_{-2} - z^3 Q_{-3} ,
\nonumber
\end{eqnarray}
which satisfy
\begin{eqnarray}
&&\cev{h_2,q_2}  \phi_{h_3,q_3}(z)   f_m(z) = 0,\;\; m\leq -1 \nonumber\\
&&\cev{h_2,q_2}  \phi_{h_3,q_3}(z)   (f_0'(z) - q_2) = 0,\\
&&\cev{h_2,q_2}  \phi_{h_3,q_3}(z)   (f_0''(z) - q_3) = 0.
\nonumber
\end{eqnarray}
Again let us put $z=1$ and drop the dependence from the generators $f_m$.
The problem now with evaluating three point functions
is that one cannot span a Verma module using only the combinations
$\{ e_0',e_0'',e_m; q_0',q_0'',q_m \}$, one also has to consider the
mode $Q_{-1}$ separately. A basis of the Verma module
is given by the states
\eq
 e_{i_1}\ldots e_{i_j} f_{k_1} \ldots f_{k_l}
(e_0')^a (e_0'')^b (f_0')^c (f_0'')^d (Q_{-1})^e
\vec{h,q}
,\;\;
\nonumber\en\eq
i_m \leq i_{m+1} \leq -1,\;
k_m \leq k_{m+1} \leq -1
{}.
\label{eq.fbasis}
\en
The three-point function of the state
\eq
\vec\psi = \sum_{a=0}
	p_a(e'_0,e''_0,e_{-1},\ldots;f'_0,f''_0,f_{-1},\ldots)(Q_{-1})^a
\vec{h,q}
\en
is clearly seen to be
\eq
\cev{h_2,q_2} \phi_{h_3,q_3}(1) \vec\psi
=
\sum_{a=0} p_a(h_2,h_3,0\ldots;q_2,q_3,0\ldots)
C_a
\label{eq.tpf}
\en
where the constants $C_a$ given by
\eq
C_a = 	\cev{h_2,q_2} \phi_{h_3,q_3}(1) (Q_{-1})^a \vec{h,q}
\en
are {\em a priori} independent coefficients.
Thus there are a countable number of fusions of generic $W_3$
representations, with parameters $\{ C_a, a\ge 0 \}$\footnote{This is
analogous to the situation in the $N=1$
superconformal algebra where in the Neveu-Schwarz sector the three
point functions $\cev{h_2} \phi_{h_2}(1) (G_{-1/2})^a \vec{h_3}$ are
independent for $a=0,1$ leading to a division of the fusion rules into
`even' and `odd'}.
We now consider the case $M_{h_1.q_1.c}$ reducible.

\subsection{3-point functions and null vectors}

As mentioned above  the structure of Verma modules over $W_3$ is not
fully understood yet, as indeed is the case even for simple Lie
algebras\footnote{I would like to thank K. de Vos for pointing this
out to me}.
There are problems arising from the fact that
$Q_0$ need not be
diagonaliseable (see ref.\ \cite{Watt1}), and the usual problem that a
submodule of a highest weight representation need not necessarily be a
sum of highest weight representations (see e.g. ref.\ \cite{BGGe1}).
It is conjectured by Bouwknegt et al.\ that these problems
are not present in the representations of the $W_3$ algebra which occur
in the unitary minimal models
which have values of $c$ satisfying
\eq
c = 2 \left( 1 - \frac{12}{m(m+1)} \right)
\,,
\label{eq.cmin}
\en
where $m$ is an integer greater than 3  (see refs.\ \cite{BMPi}
for details).

We shall initially be interested in the cases where $c$ is given by
eqn.\ (\ref{eq.cval}) but with $a^2$ irrational, and in which there are
zero, one, or two independent null vectors in the Verma module
$M_{h_1,q_2,c}$, as in cases (I),  (II) and (III) below, although
these not an exhaustive list of the possibilities.
The diagrams (I--III) show embeddings of Verma modules with the labels
being the $L_0$ eigenvalues (see e.g.\ ref.\ \cite{Watt1} for details).
We call representations with one, two and three independent null
vectors singly, doubly and completely degenerate respectively.
\vskip 4mm\par

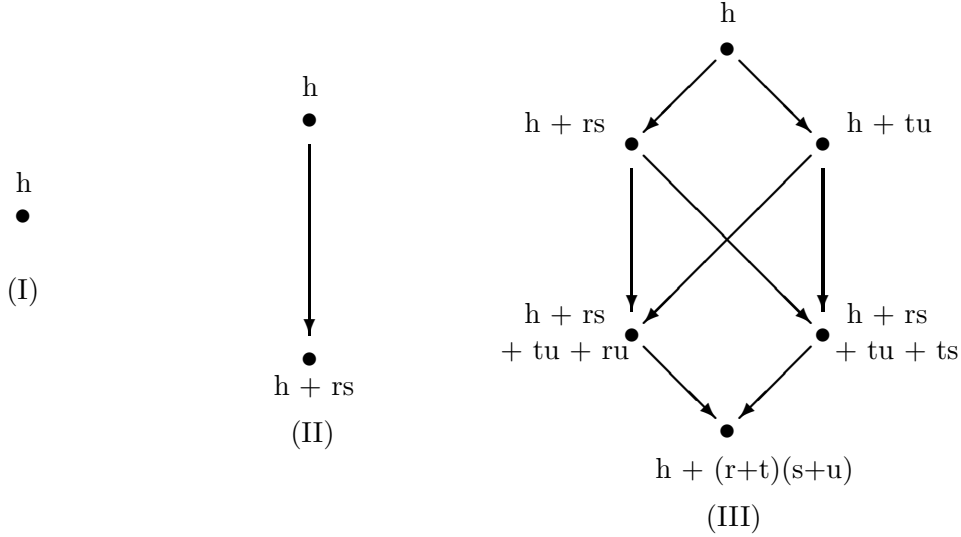
\def\twlrm{}

\centerline{\makebox[4.850in]{\rule{0in}{2.750in}
\setlength{\unitlength}{0.0125in}%
\begin{picture}(388,220)(30,560)
\thicklines
\put( 35,690){\circle*{6}}
\put( 28,655){\makebox(0,0)[lb]{\raisebox{0pt}[0pt][0pt]{\twlrm (I)}}}
\put( 32,700){\makebox(0,0)[lb]{\raisebox{0pt}[0pt][0pt]{\twlrm h}}}
\put(155,730){\circle*{6}}
\put(155,630){\circle*{6}}
\put(155,720){\vector( 0,-1){ 80}}
\put(140,615){\makebox(0,0)[lb]{\raisebox{0pt}[0pt][0pt]{\twlrm h + rs}}}
\put(152,740){\makebox(0,0)[lb]{\raisebox{0pt}[0pt][0pt]{\twlrm h}}}
\put(147,595){\makebox(0,0)[lb]{\raisebox{0pt}[0pt][0pt]{\twlrm (II)}}}
\put(290,720){\circle*{6}}
\put(330,760){\circle*{6}}
\put(370,720){\circle*{6}}
\put(290,640){\circle*{6}}
\put(370,640){\circle*{6}}
\put(330,600){\circle*{6}}
\put(325,755){\vector(-1,-1){ 30}}
\put(335,755){\vector( 1,-1){ 30}}
\put(370,710){\vector( 0,-1){ 60}}
\put(290,710){\vector( 0,-1){ 60}}
\put(295,715){\vector( 1,-1){ 70}}
\put(365,635){\vector(-1,-1){ 30}}
\put(365,715){\vector(-1,-1){ 70}}
\put(295,635){\vector( 1,-1){ 30}}
\put(245,725){\makebox(0,0)[lb]{\raisebox{0pt}[0pt][0pt]{\twlrm h + rs}}}
\put(380,725){\makebox(0,0)[lb]{\raisebox{0pt}[0pt][0pt]{\twlrm h + tu}}}
\put(235,630){\makebox(0,0)[lb]{\raisebox{0pt}[0pt][0pt]{\twlrm + tu + ru}}}
\put(245,645){\makebox(0,0)[lb]{\raisebox{0pt}[0pt][0pt]{\twlrm   h + rs}}}
\put(300,580){\makebox(0,0)[lb]{\raisebox{0pt}[0pt][0pt]{\twlrm h +
(r+t)(s+u)}}}
\put(327,771){\makebox(0,0)[lb]{\raisebox{0pt}[0pt][0pt]{\twlrm h}}}
\put(321,560){\makebox(0,0)[lb]{\raisebox{0pt}[0pt][0pt]{\twlrm (III)}}}
\put(375,630){\makebox(0,0)[lb]{\raisebox{0pt}[0pt][0pt]{\twlrm + tu + ts}}}
\put(380,645){\makebox(0,0)[lb]{\raisebox{0pt}[0pt][0pt]{\twlrm     h + rs}}}
\end{picture}
}
}

\vskip 4mm
\noindent%
Case (I). There are no highest weight vectors in $M_{h_1,q_1,c}$; this
case was dealt with in the previous section. There are a countable
number of fusions for each set of $\{h_2,q_2,h_3,q_3\}$

\vskip 2mm
\noindent%
Case (II). There is one null vector at level $rs$ in $M_{h_1,q_1,c}$.
If this is of the form%
\footnote{This is certainly the case if one of $r$ or $s$ is 1, as
explicit expressions for the null vectors have been given in \cite{BWat2}}
\eq
\vec\psi =
\sum_{a=0}^N p_a(e'_0,e''_0,e_{-1},\ldots;f'_0,f''_0,f_{-1},\ldots)(Q_{-1})^a
\vec{h_1,q_1}
\label{eq.nstate}
\en
with $p_N = 1$, then  requiring this null vector to decouple we find
using eq.\ (\ref{eq.tpf}),
\eq
C_N + \sum_{a=0}^{N-1} p_a(h_2,h_3,0\ldots;q_2,q_3,0\ldots)
C_a =0
\en
and so $C_N$ is given in terms of
$C_a, 0 \leq a <N$. Similarly by acting on the state (\ref{eq.nstate})
with $Q_{-1}$ we find that all higher $C_a$ are also determined and so
there is an $N-$dimensional space of fusions parameterised by
$\{C_0,\ldots,C_{N-1}\}$ and no restrictions on the allowed values of
$h_2,q_2,h_3,q_3$.

\vskip 2mm
\noindent%
Case (III).
If there are two independent null vectors these may give rise to a
hexagonal pattern of embeddings, and then the situation is more
complicated. The presence of two independent null
vectors in this way leads to a restriction of the values of $h_1$,
$h_2$, $q_1$ and  $q_2$ to a two-dimensional curve; that is if $h_2$
and $q_2$ are given then there is a finite set of allowed values for
$h_3$ and $q_3$.

The presence of a third independent null vector in the
$W_3$ minimal models leads to a truncation of the
fusions in case (III) to a finite set of allowed values of
$\{h_2,q_2,h_3,q_3\}$ in the same way that
the presence of a second independent null vector does in the Virasoro
minimal models.

\subsection{4-point functions}

As before, we can consider a 4-point functions as a sum over chiral
blocks, where the intermediate states run over the allowed $W_3$
algebra representations in the theory,
\eq
B((h_2,q_2),  (h_3,q_3), (h_4,q_4), (h_1,q_1); (h_5,q_5))
=
\sum_{L_{h_5,q_5}}
\cev{h_2,q_2}    \phi_{h_3,q_3}(1) \vec {\psi_\iota}
\cev{\psi_\iota} \phi_{h_4,q_4}(z) \vec{h_1,q_1}
\label{eq.cblock}
\en
If none of the $M_{h_i,q_i,c}$ are reducible than this chiral block
has no restriction on $h_5$ and $q_5$ and depends on the free parameters
$\{C_a,\tilde C_a, a \geq 0\}$ where
\eq
C_a = \cev{h_2,q_2}   \phi_{h_3,q_3}(1) (Q_{-1})^a \vec{h_5,q_5}
,\;\;
\tilde C_a = \cev{h_5,q_5}(Q_1)^a \phi_{h_4,q_4}(1) \vec{h_1,q_1}
\,.
\en
If $M_{h_1,q_1,c}$ is reducible with a single
null vector, then again the chiral block \ref{eq.cblock} has free
parameters $\{C_a, a \geq 0; \tilde C_0 \ldots \tilde C_{N-1} \}$ and
there  will be no restriction on $(h_5,q_5)$.

\noindent%
If $M_{h_1,q_1,c}$ has two independent null vectors as in case (III),
then this will determine $\tilde C_a$ up to an overall scale and
given $h_4,q_4$
will restrict $(h_5,q_5)$ to lie in a finite set, but the
three point function
\eq
\cev{h_2,q_2}   \phi_{h_3,q_3}(1) \vec \psi
\,,
\label{eq.ac}
\en
will still depend on the  infinite number of arbitrary constants $C_a$.
So we see that even if one of the fields in a four point function is
doubly degenerate, then the 4 point function still is not
restricted to lie in a finite space and so will not satisfy a
differential equation.

One way we can restrict the 4-point function further is if another of
the Verma modules, $M_{h_2,q_2,c}$ say, is doubly degenerate, as
this will again not only force the values of $(h_5,q_5)$ to lie in a
finite set, but will determine the 3-point functions (\ref{eq.ac}) up to
an overall constant.
It is only then that there will be a finite space of chiral blocks,
and that the 4-point functions can satisfy a differential equation in
$z$.
This is consistent with the results of Bajnok et al.
\cite{Bajnetal}
in which differential equations were found for some four point
functions of two doubly degenerate fields and two arbitrary fields in
the $WA_2 \equiv W_3 \equiv W(2,3)$ and $WBC_2 \equiv W(2,4)$
theories.

Another way to restrict the 4-point function is if $L_{h_1,q_1,c}$
corresponds to a completely degenerate representation in a $W_3$
minimal model. In this case $\{h_2,q_2,h_5,q_5\}$ will be restricted
to a finite set of allowed values, each of which corresponds to
another allowed minimal model representation. Since $L_{h_5,q_5,c}$ is
now restricted to a minimal model representation, this in turn will
restrict $\{h_3,q_3,h_4,q_4\}$ to a finite set of minimal model
values.

This clearly generalises to $n$-point functions. If $c$ is not a
minimal value, then there will be a finite set of chiral blocks if
$n-2$ of the fields are doubly degenerate. If $c$ is a minimal value
and one of the fields is in a completely degenerate representation then
there will only be a finite set of allowed values of $h_i,q_i,1<i\leq
n$ and a finite space of chiral blocks for each set of these values.
The fusion rules of the minimal $W_3$ representations have been given
by Frenkel et al.\ in ref.\ \cite{FKWa1} using the Verlinde formula.

\setcounter{section}{3}
\setcounter{subsection}{-1}
\section{Conclusion}

We have seen that the structure of correlation functions in theories
with $W$-algebra symmetry is rather more complicated than that in a
theory with pure conformal symmetry. In particular the constraints
arising from null vectors are weaker, and to obtain differential
equations for 4-point functions it is necessary either that at least
two of the fields are doubly degenerate, or that one of the
representations is completely degenerate,  a contrast with the case in
Virasoro theory.

I would like to thank K.\ de Vos,
M.\ D\"orrzapf,  M.\ Gaberdiel, A.\ Kent and F.\ Malikov for
stimulating conversations. I would finally like to thank
the organisers for the great efforts they have
expended to make this meeting successful and allowing me the chance to
present these results.

GMTW was supported by a research fellowship from St.\ John's
College, Cambridge.
PB was supported by DOE of USA grant Number DEFG02-90-ER-40560 and NSF grant
PHY900036.

\end{document}